\documentclass[conference]{IEEEtran}
\IEEEoverridecommandlockouts
\PassOptionsToPackage{table, x11names, dvipsnames, svgnames}{xcolor}
\usepackage{cite}
\usepackage{bbm}
\usepackage{amsmath,amssymb,amsfonts}
\usepackage{mathtools}
\usepackage{algorithmic}
\usepackage{graphicx}
\usepackage{textcomp}
\usepackage{xcolor}
\usepackage[english]{babel}
\usepackage[acronym,nonumberlist]{glossaries}
\usepackage{glossaries-prefix}
\usepackage{pgfplots}
\pgfplotsset{compat=1.17}
\usepgfplotslibrary{fillbetween}
\usepackage{amssymb}
\usepackage{import}
\usepackage{bm}
\usepackage{subfig}
\usepackage{siunitx}
\usepackage{multirow}
\usepackage[]{yquant}
\usepackage{hyperref}
\usepackage{braket} 
\hypersetup{%
linktocpage=true, 
colorlinks=false,
pdfborder={0 0 0}, 
breaklinks=true, pdfpagemode=UseNone, pageanchor=true, pdfpagemode=UseOutlines,%
plainpages=false, bookmarksnumbered, bookmarksopen=true, bookmarksopenlevel=1,%
hypertexnames=true, pdfhighlight=/O,
pdftitle={CVRP and circuit cutting},%
pdfauthor={Christian Ufrecht},%
pdfsubject={QuaST},%
pdfkeywords={},%
pdfcreator={pdfLaTeX},%
pdfproducer={LaTeX with hyperref}%
}

\usepackage[inline]{enumitem}

\usepackage[capitalise]{cleveref}

\usepackage{xcolor}

\begin{document}



\title{Improving Quantum and Classical Decomposition Methods for Vehicle Routing

\thanks{
* These authors contributed equally (name order randomised). \\
FW thanks Frauke Liers for many fruitful discussions.\\
The research was funded by the project QuaST, supported by the Federal Ministry for Economic Affairs and Climate Action on the basis of a decision by the German Bundestag.\\
email address for correspondence:\\
christian.ufrecht@iis.fraunhofer.de}
}


\author{\IEEEauthorblockN{Laura S. Herzog\IEEEauthorrefmark{1}\IEEEauthorrefmark{2}, Friedrich Wagner\IEEEauthorrefmark{1}\IEEEauthorrefmark{2}\IEEEauthorrefmark{3},
Christian Ufrecht\IEEEauthorrefmark{1}\IEEEauthorrefmark{2},
Lilly Palackal\IEEEauthorrefmark{5} \\
Axel Plinge\IEEEauthorrefmark{2}, Christopher Mutschler\IEEEauthorrefmark{2
}, and Daniel D. Scherer\IEEEauthorrefmark{2}}
\IEEEauthorblockA{\IEEEauthorrefmark{2}
Fraunhofer Institute for Integrated Circuits IIS}
\IEEEauthorblockA{\IEEEauthorrefmark{5}Infineon Technologies AG}
\IEEEauthorblockA{\IEEEauthorrefmark{3}University of Erlangen-Nuremberg, Department of Data Science}
}

\maketitle

\begin{abstract}
Quantum computing is a promising technology to address combinatorial optimization problems, for example via the quantum approximate optimization algorithm (QAOA).
Its potential, however, hinges on scaling toy problems to sizes relevant for industry.
In this study, we address this challenge by an elaborate combination of two decomposition methods, namely graph shrinking and circuit cutting.
Graph shrinking reduces the problem size before encoding into QAOA circuits, while circuit cutting decomposes quantum circuits into fragments for execution on medium-scale quantum computers.
Our shrinking method adaptively reduces the problem such that the resulting QAOA circuits are particularly well-suited for circuit cutting.
Moreover, we integrate two cutting techniques which allows us to run the resulting circuit fragments sequentially on the same device.
We demonstrate the utility of our method by successfully applying it to the archetypical traveling salesperson problem (TSP)
which often occurs as a sub-problem in practically relevant vehicle routing applications.
For a TSP with seven cities, we are able to retrieve an optimum solution by consecutively running two 7-qubit QAOA circuits.
Without decomposition methods, we would require five times as many qubits.
Our results offer insights into the performance of algorithms for combinatorial optimization problems within the constraints of current quantum technology.
\end{abstract}


\section{Introduction}
The semiconductor supply of a country is of political importance.
Hence, resilient supply chain planning is crucial, especially in times of crises.
Among other challenges, this requires solving a range of NP-hard optimization problems in large scale on a daily basis.
In order to meet customers' demands, production routes distributed among several fabrication facilities are considered.
Many of these routing problems can be modelled as a capacitated vehicle routing problem (CVRP)~\cite{Toth2002}.
\begin{figure}[h]
\centering
\includegraphics[width=246pt]{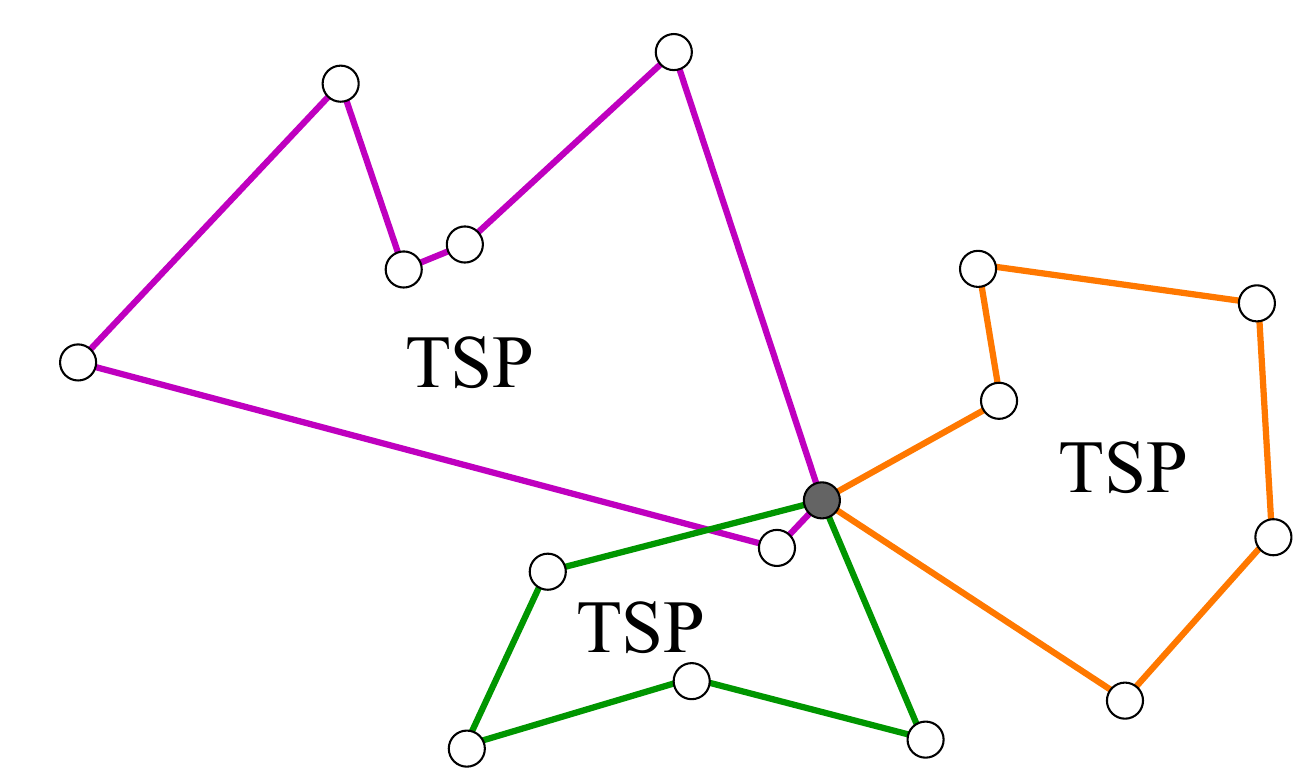}
\caption{An example for the capacitated vehicle routing problem. 15 customers (white circles) need to be served by three trucks of limited capacity, initially located at the depot (filled circle). A heuristic decomposition method first clusters the customers into groups which can be served by a single truck. Then, a Traveling Salesperson Problem is solved for each cluster individually (colored lines).} 
\label{fig:cvrp_ex}
\end{figure}
As nowadays classical algorithms take very long to solve this problem optimally, supply chain planners work with a mix of solvers and heuristics which enable good solutions in decent time.
However, there is still room for improvement with respect to performance or time to solution. 
\begin{figure*}
	\centering
	\subfloat{\includegraphics[width=1\textwidth]{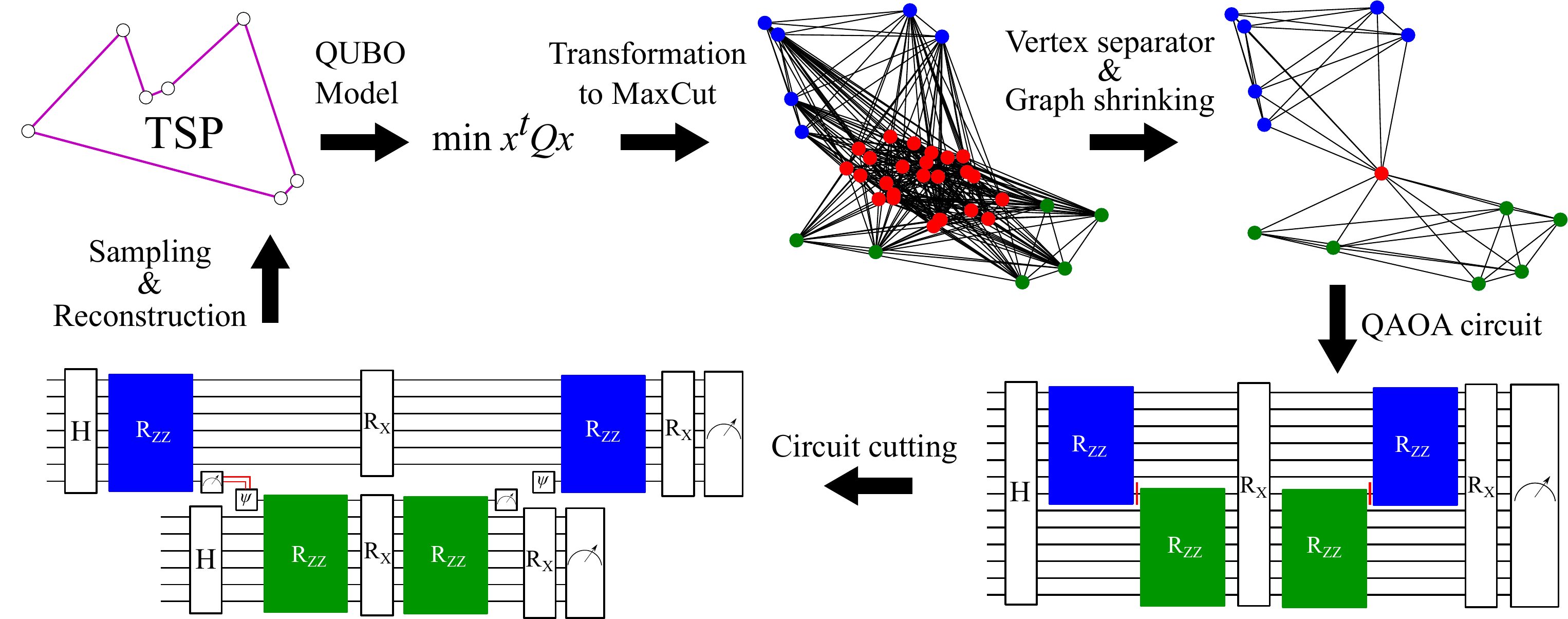}}
	\caption{Workflow. Starting from the TSP instance with 7 nodes, we first bring the problem in QUBO form with dimension 36. Any QUBO problem can be reformulated as a weighted MaxCut instance by introducing one extra node, which is done in the next step. We are now in the position to apply our heuristic graph partitioning algorithm to search for a balanced vertex separator, which separates the graph into the green and blue node subset which are connected only via the vertex separator consisting of the nodes marked in red. The subsequent application of shrinking reduces the graph to two equally sized subgraphs with six nodes each, connected by a single-node vertex separator marked in red. This form leads to QAOA circuits that can be decomposed by a single wire cut per layer shown by the red bars in the circuit. By leveraging a combination of circuit-cutting methods that require one-way classical communication only, we are able to train and evaluate the QAOA on a quantum computer with 7 qubits. The flow of classical communication is shown by the red connections between measurement and state-initialization symbol.
    Sampling bit strings from the trained circuit, we find the optimal solutions to the original TSP instance after reverting the shrinking process.
	}
	\label{fig:workflow}
\end{figure*}
Quantum computing~\cite{Nielsen2010} has emerged as a promising technology for solving such combinatorial optimization problems which are notoriously difficult for classical computers.
However, most quantum algorithms for this class of problems are heuristic in nature without rigorous guarantees~\cite{abbas2023quantum}.
Consequently, potential benefits from the use of quantum computers might only become present as the problem size reaches practically relevant scales,
a significant challenge for quantum computers available today and in the near term.
Furthermore, the impact of noise on quantum computations in the noisy intermediate-scale quantum era (NISQ)~\cite{Preskill2018} severely deteriorates the quality of computations.

In this paper, we explore the combination of two methods aimed at reducing problem sizes to make quantum computations more feasible.
The first method, known as graph shrinking \cite{Liers_2011,Wagner2023}, is a classical technique for reducing the size of a problem before encoding it into a quantum circuit.
The second method is known as circuit cutting \cite{Hofmann2009, Peng2020} and is a pure quantum method.
It can be used to partition quantum circuits in a way such that the resulting independent fragments can be evaluated on smaller quantum hardware.
In our study we address two important questions.
First, we investigate how graph shrinking in combination with circuit cutting impacts the quality of solutions.
Second, we formally discuss the application of circuit cutting to sampling tasks. This is an important problem to address since after training the result of the QAOA is obtained by sampling bit strings.
Our evaluation is conducted for problems describing the optimization of semiconductor supply chains.
By examining the CVRP as an exemplary combinatorial optimization problem, we aim to demonstrate the effectiveness of our approach and to highlight the trade-offs involved in such methods.

\section{Problem definition}
In this section we first define the CVRP more formally.
We then introduce the transformation into a Quadratic Unconstrained Binary Optimization (QUBO) problem, which is required as an input for the quantum approximate optimization algorithm.
Finally, the technique of quantum circuit cutting is introduced. The workflow introduced in this work is shown in \cref{fig:workflow}.

\subsection{The CVRP}
Mathematical models for complex semiconductor supply chains include a range of optimization problems, such as routing problems.
One of the most renowned problems is the CVRP~\cite{Toth2002}.
The CVRP is defined on a complete graph with weighted edges and nodes.
Node $0$ represents a depot and each node $i\geq 1$ corresponds to a customer with demand (node weight) $\delta_i$.
Given a truck with capacity $C$, the task is to find the shortest route, i.e. the route with the least edge weight,
visiting all customers and returning to the depot while not overstepping the capacity of the truck.
An illustrative example is waste collection.
Here, each node represents a waste bin with filling level $\delta_i$.
The truck can transport a maximum of $C$ units of waste and the depot is a dumpsite where the truck can unload the waste and start the next round with full capacity.
As the CVRP has been proven to be NP-hard, it is not expected that there exists a classical algorithm that can efficiently determine the shortest route for general large instances.
While quantum algorithms are not expected to efficiently solve NP-hard problems either,
they promise to find better solutions or faster approximations~\cite{abbas2023quantum}.
Recent studies derived Quadratic Unconstrained Binary Optimization (QUBO) formulations for the CVRP.
In the work of Feld et al.~\cite{Feld2019} a QUBO formulation for the CVRP is given which is slightly modified in Ref.~\cite{Palackal2023}, requiring fewer binary variables and thus fewer qubits.
Nevertheless, even for small instances, the number of qubits required still far exceeds the limitations of today's quantum hardware.
For instance, following the QUBO formulation given in Ref.~\cite{Palackal2023}, the instance depicted in \cref{fig:cvrp_ex} requires hundreds of qubits.
Therefore, in order to solve CVRP instances on smaller quantum computers, suitable problem decomposition and circuit cutting methods, as introduced in this work, are essential.

An established class of heuristics for the CVRP are cluster-first route-second approaches, compare e.g.~\cite{Toth2002} for a comprehensive survey.
In such a heuristic, the customers are first grouped into clusters such that each cluster can be served by a single vehicle.
In a second step, the shortest route is determined in each cluster, see \cref{fig:cvrp_ex} for an example.
Finding the shortest route in a cluster amounts to solving the well-known traveling salesperson problem (TSP).
Given a complete graph $G=(V,E)$ with distances $d_{ij}>0$ for $ij\in E$, the TSP asks for a Hamilton cycle in $G$ of minimal length.
The study in Ref.~\cite{Feld2019} proposes a promising quantum-classical hybrid approach to the CVRP
which performs clustering classically and subsequently solves the TSPs via a quantum algorithm.
In this work, we follow this approach and develop quantum solution methods for the TSP,
keeping in mind that the TSP might originate from a clustering decomposition of a CVRP problem.

\subsection{QUBO and MaxCut}
Today, most implementations of quantum algorithms for combinatorial optimization are restricted to QUBO problems,
that is, problems of the form
\begin{align} 
    \min_{x\in \{0,1\}^n} \sum_{i,j}q_{ij}x_ix_j\,, \label{eq:qubo}\
\end{align}
where $q_{ij} \in \mathbb{R}$.
Thus, solving a given application problem such as the TSP via QAOA requires a reformulation of the problem as a QUBO.
Here, we work with the standard QUBO model for the TSP~\cite{Lucas_2014}.
For a given TSP instance on $N$ vertices, the model requires $n=(N-1)^2$ variables.

It is well-known that any QUBO problem can be translated into an equivalent maximum cut problem (MaxCut)~\cite{Hammer_1965,Barahona_1989,DESIMONE_1990,Juenger2021}.
Given a weighted graph $G=(V,E)$ with edge weights $w_{e}, e \in E$, MaxCut asks for a partition of the vertices into two subsets such that the weight of connecting edges is maximized.
A QUBO problem on $n$ variables can be transformed into an equivalent MaxCut problem on $n+1$ vertices.
Transforming the QUBO into a MaxCut problem allows us to reduce the problem size via graph shrinking which we explain in more detail in \cref{sec:shrinking}.
Details on the TSP QUBO model and on the transformation into MaxCut can be found in \cref{QUBO Model for the TSP}.

\subsection{Circuit cutting}
Circuit cutting is a technique for the decomposition of large quantum circuits into smaller parts, with the additional constraint that the resulting fragments can be executed on quantum hardware.
More formally, circuit cutting decomposes a quantum channel $\mathcal{W}$ into a sum 
\begin{equation}
\label{decomposition}
    \mathcal{W}=\sum_i a_i \mathcal{F}_i \,,
\end{equation}
where $a_i$ are real numbers.
There are two primary circuit cutting methods, known as gate cutting and wire cutting.
In gate cutting~\cite{Hofmann2009, Mitarai2021b, Piveteau2022_circuitcut, Mitarai2021, Ufrecht2023,Ufrecht2023b, Schmitt2024, Harrow2024}, $\mathcal{W}$ is a non-local unitary channel and $\mathcal{F}_i$ are local unitary or measurement channels.
On the other hand, wire cutting~\cite{Brenner2023, Peng2020, Pednault2023, Lowe2022, Harada2023, Uchehara2022} refers to a strategy for severing qubit wires, where $\mathcal{W}$ is the identity channel describing empty qubit lines and $\mathcal{F}_i$ are measure-and-prepare channels. 

As hardware quality proceeds, circuit cutting could be combined with distributed quantum computing \cite{Vazquez2024, Devoret2013,  Monroe2013} and  virtual state distillation \cite{Yuan2023,Bechtold2024}.

For the application in this work, we will focus on wire cutting.

All circuit cutting methods incur an exponential sampling overhead compared to the uncut circuit.
This factor of more samples required to estimate the expectation value of an observable to a given accuracy with the cut circuit is in $\mathcal{O}(\kappa^{2K})$.
Here, $K$ is the number of gate or wire cuts and
\begin{equation}
\label{definition_kappa}
   \kappa=\sum_i|a_i|
\end{equation}
is determined by the specific decomposition employed.
Consequently, an essential objective for circuit cutting research is the identification of a decomposition with a minimal value of $\kappa$.
Typically, one tries to cut all connections between two predefined partitions such that the circuit disintegrates into two independent parts which can be executed on smaller quantum devices.
Of course the resulting partitions can be further decomposed if necessary.

To obtain an estimate for the expectation value of an observable $O$ from a quantum computer, one typically defines a postprocessing function $f(s)$ for bit string $s$ measured with probability $p(s)$ such that $\sum_s p(s) f(s)=\langle O\rangle$.
In case of a cut circuit, we replace the channel $\mathcal{W}$ by the sum given in \cref{decomposition}.
Consequently,
\begin{align}
\label{quasi-prob-sampling}
    \langle O\rangle &= \sum_{i,s} a_i\, p(s|i)f(s)\\
    \label{quasi-prob-sampling1}
    &=\kappa \sum_{i,s} p(i) \, p(s|i)\,\mathrm{sign}(a_i)f(s) \,,
\end{align}
where $p(s|i)$ denotes the probability of observing $s$ if $\mathcal{W}$ is replaced by the $i$th channel $\mathcal{F}_i$.
In the second equality we defined the probability
\begin{equation}
\label{equ-pi}
p(i)=|a_i|/\kappa\,.
\end{equation}
The expression in \cref{quasi-prob-sampling1} can be simulated by quasi-probability sampling.
At each shot, a channel $\mathcal{F}_i$ is sampled according to $p(i)$.
The circuits for both partitions are executed and the two bit strings measured are concatenated.
Then, the concatenated bit string is passed to the postprocessing function which is subsequently weighted with $\kappa\,\mathrm{sign}(a_i)$.
The mean over many repetitions finally produces an unbiased estimate for $\langle O\rangle$. 
For wire cutting, $\mathcal{F}_i$ are measure and prepare channels which are randomly applied at each experimental run.
Each measure and prepare channel defines a measurement basis and a state the qubit is re-initialized in after the measurement.
This process effectively corresponds to cutting the wire.
We emphasize the importance of the sign of $a_i$ in \cref{quasi-prob-sampling1}, therefore the term quasi-probability sampling.
We note, as further detailed in~\cref{sec:choice_cutting}, that in many cases it is possible to calculate how many times each circuit has to be executed prior to the experiment, allowing evaluation in batches which reduces latency times. 
Due to the exponential growth of the sampling overhead, a naive application of circuit cutting typically becomes impractical after only cutting a few gates or wires. 
As described in the next section, this problem can be overcome by an anticipatory application of graph shrinking which ensures that the number of cuts later needed to partition the circuit is minimal.
Furthermore, we carefully select a combination of different wire cutting methods to further reduce the sampling overhead.

\section{Methodology}
In this section, we describe the proposed workflow for solving large combinatorial optimization problems on quantum processors of limited size in detail.
We illustrate this workflow by considering the TSP as an archetypical combinatorial optimization problem but we emphasize that our method is applicable to any input problem.
Similarly, we restrict our analysis to depth-two QAOA, but adapting our method to higher-depth QAOA is straightforward.

The overall concept is summarized in \cref{fig:workflow}.
First, the application problem, in our case the TSP, is modeled as a QUBO problem.
Then, this QUBO is transformed into a MaxCut instance.
Subsequently, this MaxCut instance is reduced in size such that a low-cardinality vertex separator exists \cite{Lowe2022}.
We construct the corresponding QAOA circuit and apply wire cutting at the low-cardinality vertex separator.
Finally, both partitions are run sequentially on quantum hardware and solutions to the TSP are reconstructed from the retrieved samples.
We now describe the individual steps in detail.

\subsection{Graph Shrinking}\label{sec:shrinking}
We adopt the graph shrinking technique developed in previous work~\cite{Wagner2023}.
First, we transform the QUBO problem into a MaxCut instance $G=(V,E)$.
This allows us to use established linear programming techniques for MaxCut.
In particular, we use a common integer linear programming formulation of MaxCut,
which introduces a binary variable $x_e\in \{0,1\}$ for all edges $e\in E$.
A value of $x_e=1$ ($x_e=0$) indicates that edge $e$ connects vertices in opposite (equal) partitions~\cite{Barahona_1989}.
A detailed discussion of the integer linear model is given in \cref{app:cycle_rel}

Although the MaxCut problem is NP-hard,
optimizing the linear programming (LP) relaxation of the integer model can be done in polynomial time.
The LP relaxation is obtained from the integer model by allowing the variables to take continuous variables $x_e\in [0,1]$.
A key ingredient of our algorithm is that a solution to the LP relaxation
can be computed efficiently in practice.

We use an optimum solution of the LP relaxation to reduce the size of the
MaxCut instance such that it is well suited for circuit cutting.
To this end, we chose an edge $e \in E$ whose corresponding value $x_e$ in the relaxation solution is close to either 0 or 1
and enforce the corresponding vertices to lie in equal or opposite partitions, i.e.~we fix the value of $x_e$ to either 0 or 1.
The rationale for this is that variables which are close to integer in a relaxation solution often take the corresponding integer value in an optimum MaxCut solution.

\begin{figure}
	\centering
	\subfloat[]{\includegraphics[height=2.5cm]{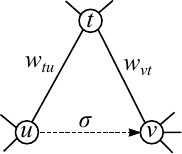}\label{fig:skizze_vertex_indent_vor}}\hspace{2cm}
	\subfloat[]{\includegraphics[height=2.5cm]{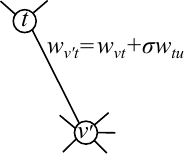}\label{fig:skizze_vertex_indent_nach}}
	\caption{Sketch for vertex shrinking.
		(a): Vertices $u$ and $v$ are to be identified, where
		$\sigma\in \{-1,1\}$ defines whether $u$ and $v$ lie in equal ($\sigma = 1$) or opposite ($\sigma = -1$) partitions.
		Vertex $t$ is a neighbor of $u$.
		(b): In the shrunk MaxCut instance, vertex $v'$ is a super-vertex containing $u$ and $v$ with adjusted edge weights.
	}
	\label{fig:shrinking}
\end{figure}

Then, solving the MaxCut problem with the fixed value for $x_e$ is equivalent to solving a MaxCut problem where the number of vertices is reduced by one.
This reduced MaxCut problem is constructed by contracting the vertices incident to $e$ into a single vertex and adjusting the weights appropriately as illustrated in \cref{fig:shrinking}.
We refer to such a vertex resulting from contracting an edge as a super-vertex.
Any MaxCut solution for the shrunk graph can be translated into a MaxCut solution for the original graph by reverting the shrinking procedure.
We iterate this procedure until a desired graph size is reached.
This process is called \emph{graph shrinking}~\cite{Liers_2011, Bonato_2014}.
The number of shrinking steps, i.e., edge contractions should be as small as possible since shrinking can incur errors if no optimum solution exists obeying the imposed variable fixings.

The goal of the graph shrinking procedure is to arrive at a shrunk graph which has a vertex separator of low cardinality.
A vertex separator of a graph is a subset of vertices which, if removed, decomposes the graph into two or more connected components~\cite{Rosenberg2002}.
Thus, the vertex set $V'=V\setminus C$ of the graph with removed separator $C\subset V$ can be partitioned into a disjoint union $V'=A\dot{\cup}B$ such that no edge connects $A$ and $B$, see \cref{fig:workflow} for an illustration.
In the corresponding QAOA circuit, the cardinality of the separator equals the number of wire cuts necessary to partition a single layer of the QAOA circuit~\cite{Lowe2022}.
As the sampling overhead grows exponentially with the number of cuts, we aim at finding a vertex separator of small cardinality.
In our experiments, we shrink such that the resulting graph has a vertex separator of cardinality one, but we remark that separators of any desired cardinality can be produced by our method.

As an additional constraint, the resulting partitions $A$ and $B$ should be of roughly equal size such that the number of required qubits $\max\{|A|,|B|\}$ is minimized.
A separator which partitions a graph into equally sized subsets is called a \emph{balanced} vertex separator~\cite{Feige2008,Althoby2020}.

To satisfy these requirements, we first find a minimal cardinality vertex separator in the original graph such that the resulting partitions $A$ and $B$ have roughly equal size.
Then, we shrink the vertex separator to a single super-vertex which then forms a vertex separator in the shrunk graph.

Computing a  minimum cardinality balanced vertex separator is NP-hard~\cite{BUI1992}.
However, integer programming techniques allow the solution of instances with hundreds of nodes in less than one second~\cite{Althoby2020}.
In order to find a minimum cardinality balanced vertex separator, we adopt the integer programming model from Ref.~\cite{Althoby2020}.
To this end, we first define $\beta\in \mathbb{Z}_{\geq 0}$ to be the maximum allowed difference in cardinality of the resulting partitions.
Furthermore, we introduce binary variables $x_v,y_v \in \{0,1\}$ for all $v \in V$.
Variable $x_v$ ($y_v$) indicates that vertex $v$ is in partition $A$ ($B$).
Then, the problem of finding a minimum cardinality balanced vertex separator problem can be modeled as
\begin{subequations}
	\begin{align} 
		\max_{x,y} &\sum_{v \in V} x_v+y_v\label{sep_obj}\\
		\text{s.t.\quad}&x_{u}+y_{v}  	\leq 1\quad  \forall uv \in E\label{sep_edge}\\
		&x_{v}+y_{u}  	\leq 1\quad  \forall uv \in E\label{sep_edge_prime}\\
        &x_{v}+y_{v}  	\leq 1\quad  \forall v \in V\label{sep_vertex}\\
        &\sum_{v \in V} x_{v}-y_{v} \leq \beta\label{sep_card1}\\
        &-\sum_{v \in V} x_{v}-y_{v} \leq \beta\label{sep_card2}\\
  	 & x_v,y_v \in \{0,1\} \quad\forall v \in V\ \nonumber
	\end{align}
\end{subequations}
The objective \eqref{sep_obj} minimizes the cardinality $|C|=|V|-(|A|+|B|)$ of the separator.
Constraints~\eqref{sep_edge} and \eqref{sep_edge_prime} ensure that no edge connects $A$ and $B$.
Constraint \eqref{sep_vertex} ensures that no vertex belongs to both $A$ and $B$.
Finally, Constraints \eqref{sep_card1} and \eqref{sep_card2} ensure that $||A|-|B||\leq \beta$.

Having identified a separator, we iteratively shrink edges in the separator until a single vertex is left, compare~\cref{fig:workflow}.
At each shrinking step, we calculate an optimum LP relaxation solution and shrink the edge in the separator whose value is closest to an integer.
After shrinking, we construct the depth-two QAOA circuit corresponding to the shrunk MaxCut instance.
This circuit is now well-suited for the application of wire cutting on the qubit corresponding to the vertex separator (red vertex and qubit line in~\cref{fig:workflow}).

\subsection{Choice of the circuit-cutting method}
\label{sec:choice_cutting}
With the result of the shrinking approach, we are now in the position to construct the QAOA quantum circuits and select the circuit cutting technique.
Circuit cutting has been applied to QAOA circuits before. In Ref. \cite{Bechtold2023a} the authors investigate the sampling overhead from cutting two-qubit rotation gates in the circuits. Refs. \cite{Saleem2021,Lowe2022} employ graph-partitioning techniques similar to the present work. Our work adds two powerful components to reduce problem size and sampling overhead from cutting: graph shrinking and the combination of two circuit-cutting methods as described in this section.

Several methods for wire cutting with reduced sampling overhead have been proposed, following the initial work by Peng et al.~\cite{Peng2020}.
These ideas will be briefly outlined in the following to motivate our approach.
The method by Peng et al. \cite{Peng2020} comes with $\kappa=4$, resulting in a sampling overhead of $16^K$ for $K$ cuts.
It was later discovered that this overhead is optimal as long as no classical communication is allowed~\cite{Brenner2023, Harada2023}.
If a wire cutting scheme requires classical communication, the state that is to be initialized after the cut depends on the measurement outcome at the cut.
Interestingly, it was shown that with classical communication, optimality is achieved with $\kappa=3$, leading to a sampling overhead of $9^K$~\cite{Harada2023,Brenner2023}. 
However, utilizing classical communication, the first cut in the bottom-left circuit in \cref{fig:workflow} would require communication from the upper to the lower and the second cut from the lower to the upper partition. 
Evaluating a circuit cutting protocol that involves this type of two-way classical communication necessitates two quantum computers operating in parallel with real-time communication links.
Additionally, since we cannot determine in advance how frequently each measurement result will occur, we cannot predetermine the number of shots required for each combination of measurement basis and new input state at the cuts.
Consequently, the measure and prepare channels have to be selected on a shot-to-shot basis, introducing large latency times.
To strike a balance between reducing sampling overhead and ensuring experimental practicability, we choose to alternate between the methods proposed by Harada et al.~\cite{Harada2023} (classical communication) and Peng et al.~\cite{Peng2020} (no classical communication).
As a result, we only require one-way classical communication even for an arbitrary number of QAOA layers.
Thus, we are able to first execute the circuits on the upper partition, whose measurement outcomes determine the number of times each circuit on the lower partition needs to be executed in the second step.
The explicit formulas are stated in \cref{app-combiningTwoCircuitCuttingTechniques}. 
For an even number of layers, we therefore arrive at a sampling overhead of $12^p$ where $p$ is the number of QAOA layers.
We remark that $\kappa$ can be further reduced 
through a joint decomposition of all cuts~\cite{Brenner2023}, albeit at the expense of additional ancilla qubits.
Furthermore, our method can be applied to circuits derived from graphs with arbitrary vertex separator cardinality.
In such cases, more cuts per layer are necessary which further increases the sampling overhead although some reduction is possible through joint decomposition schemes~\cite{Harada2023, Brenner2023, Lowe2022}.

\subsection{Circuit cutting for sampling tasks}\label{sampling-tasks}
After parameter optimization based on expectation value maximization, QAOA becomes a sampling task. 
As shown in \cref{quasi-prob-sampling} and \cref{quasi-prob-sampling1}, quasi-probability sampling can be employed to obtain unbiased estimates of expectation values, even allowing to simulate exactly the effect of gates entangling the partitions solely in terms of local gates and measurements. 
However, these guarantees are meaningless if the task shifts towards sampling bit strings from the original circuit and we aim to describe the bit string distribution generated by the cut circuit.
Nevertheless, some statements are still possible~\cite{Lowe2022} that will be presented here in generalized more rigorous form.
At each experimental shot the circuit is executed with $\mathcal{F}_i$ instead of $\mathcal{W}$ with probability $p(i)$.
The resulting concatenated bit strings are then distributed according to 
\begin{align}
\label{eq:sampling_task1}
    \tilde{p}(s)&=\sum_i p(i)\, p(s|i)\\
    \label{sampling_task2}
    &=\frac{1}{\kappa}\sum_i|a_i|\,p(s|i)\geq  \frac{1}{\kappa} \Big|\sum_i a_i p(s|i)\Big|\\
    \label{sampling_task3}
    &=\frac{1}{\kappa} p(s)\,.
\end{align}
In \cref{sampling_task2} we first used \cref{equ-pi} and then applied the triangle inequality.
In the final step to \cref{sampling_task3} we used $\sum_i a_i\, p(s|i)=p(s)$ which follows from the definition in \cref{quasi-prob-sampling}.
Note the crucial difference between \cref{eq:sampling_task1} and \cref{quasi-prob-sampling}. In the sampling task we cannot account for the sign of $a_i$ since no expectation is evaluated. This is the reason why the probability distribution of bit strings changes from $p(s)$ to $\tilde{p}(s)$ when sampling from the cut circuit. 
The statement $\tilde{p}(s)\geq p(s)/\kappa$ guarantees that any bit string with a non-zero probability in the uncut circuit, will also have a non-zero probability after cutting the circuit.
For a given number of samples $N$ for the original circuit, we define $\tilde{N}$ as the number of samples necessary in the cut scenario such that the probability to observe a given bit string at least once is of the same size.
In \cref{Circuit cutting and sampling} we argue that 
\begin{equation}
\label{overhead_sampling_task}
    \tilde{N}\approx\kappa N
\end{equation}
holds.
Surprisingly, unlike in the case of expectation value estimation, where the sampling overhead is $\mathcal{O}(\kappa^2)$, for sampling tasks, the overhead is linear in $\kappa$.

Having described the overall workflow,
we are now in the position to evaluate our method
on an exemplary optimization problem.

\section{Results}
In this section we present numerical results showcasing the applicability of the proposed methodology.
Our evaluation addresses two main research questions.
\begin{enumerate}
    \item Is an optimum solution to the application problem recovered after applying graph shrinking and circuit cutting?
    \item How does circuit cutting influence the probability distribution over bit strings, i.e.,~solutions?
\end{enumerate}

All implementations are written in Python.
The (integer) linear models are solved via Gurobi~\cite{Gurobi}.
For graph operations and quantum circuit simulation, we use the packages Networkx~\cite{hagbger2008networkx} and Qiskit~\cite{Qiskit}, respectively.

As an exemplary optimization problem, we consider the seven-city TSP shown at the top left of \cref{fig:workflow}.
It is generated by randomly placing seven cities in the euclidean plane.
Consequently, the resulting QUBO has 36 variables which is transformend into a MaxCut instance with 37 vertices, shown in \cref{fig:workflow}.
We calculate a minimum cardinality balanced vertex separator with $\beta=0$ in \cref{sep_card1,sep_card2}.
The resulting separator has 25 vertices, marked in red in \cref{fig:workflow}.
We then solve the LP relaxation of the MaxCut integer model and shrink the edge in the separator whose corresponding variable value is closest to an integer.
We repeat this process until a single vertex of the separator is left as shown in \cref{fig:workflow}.
As already mentioned in \cref{sec:shrinking}, shrinking can in principle incur errors in the sense that an optimum solution to the shrunk MaxCut instance
does not give rise to an optimum solution to the original instance.
However, this is not the case here: although shrinking 24 vertices, which amounts to a reduction of 65 \%, an optimum TSP solution can still be recovered from the shrunk problem.

After shrinking we construct the corresponding two-layer QAOA circuit.
Due to the anticipatory shrinking, the circuit can be separated into two independent, equally-sized sub-circuits by applying two wire cuts, marked in red.
As described in \cref{sec:choice_cutting}, two different circuit cutting methods are applied consecutively. They are chosen such that only one-way classical communication is required (red double-wires in \cref{fig:workflow}).
We first apply the method of Harada et al.~followed by the method of Peng et al.
This determines the flow of classical communication from the upper to the lower partition as illustrated in \cref{fig:workflow}.

\begin{figure}[ht]
	\centering
	\subfloat{\includegraphics[width=252pt]{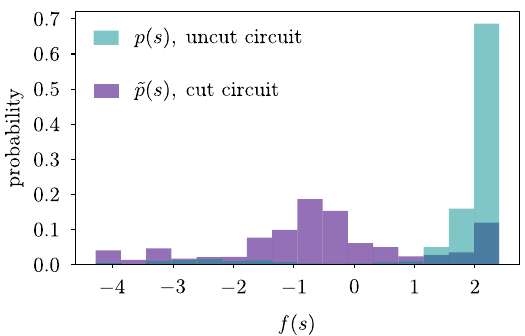}}\\
	\caption{Comparison of the bit string distributions $p(s)$ and $\tilde{p}(s)$ after training the QAOA circuit. While the teal distribution resulting from the uncut circuit is peaked close to the maximum, the purple distribution is shifted far to the left. Note, the purple peak at the maximal values of $f(s)$ which is guaranteed by the inequality between \cref{eq:sampling_task1} and \cref{sampling_task3}.
	}
	\label{fig:probability_plot}
\end{figure}

We optimize the QAOA parameters $(\gamma_1,\beta_1,\gamma_2,\beta_2)$ via maximizing the expectation value $\langle O \rangle$,
employing the continuous optimizer COBYLA~\cite{Powell1994ADS}.
After parameter optimization, the probability distribution $p(s)$ over bit strings for the uncut circuit is computed via statevector simulation to simulate the arising probability distributions exactly.
Similarly, we calculate $\tilde{p}(s)$ according to \cref{eq:sampling_task1}, describing the distribution when sampling bit strings from the cut circuit.
The explicit form of this equation for the particular problem circuits discussed here can be found in \cref{app-combiningTwoCircuitCuttingTechniques}, \cref{eq:p_cut}.
The two probability distributions $p$ and $\tilde{p}$ are shown in \cref{fig:probability_plot}.
After parameter optimization, the probability distribution generated by the uncut circuit (teal) peaks close to the optimal value of $o=2.405$
with an expectation value of $\langle O \rangle \approx 1.762$.
In contrast, cutting severely broadens the distribution $\tilde{p}(s)$ (purple) and worsens the expectation value to $-0.507$ calculated with respect to $\tilde{p}(s)$.

However, we note a peak close to the optimal value of $f(s)$ which guarantees that even in the cut scenario, optimal bit strings are sampled frequently. This peak is guaranteed by the inequality between \cref{eq:sampling_task1} and \cref{sampling_task3}.
Since the graph shrinking did not incur any errors, optimal bit strings correspond to optimal TSP solutions.

\section{Discussion}
In this work, we combined two decomposition methods to reduce the size of quantum circuits, enabling the application to medium-sized problems. 
Starting from the CVRP, we decomposed the problem into several TSP instances via clustering.
Once brought into QUBO form, graph shrinking further reduced the problem size and made it suitable for the quantum circuit decomposition approach of wire cutting.
Here, we integrated two different cutting methods to minimize sampling overhead. 

From our work, the inherent challenge of circuit cutting becomes evident.
A lower bound on the minimal possible sampling overhead is determined by the strength of entanglement between the circuit partitions \cite{Piveteau2022_circuitcut}.
Higher entanglement generally leads to greater sampling overhead.
It has been observed that this poses serious practical limitations for using circuit cutting in e.g.~Trotterized time evolution where the states are known to be highly entangled~\cite{Gentinetta2024}.
We acknowledge that a combinatorial optimization problem on a graph with a small-cardinality vertex separator as used in the present work can be decomposed and solved independently by classical means with far less computational overhead compared to the quantum runtime overhead from circuit cutting. 
However, we emphasize that our approach serves as a first step towards combining classical and quantum size-reduction methods.
Indeed, the QAOA target Hamiltonian is classical and has a non-entangled ground state.
Consequently, we might anticipate a much more efficient inter-layer joint cutting approach.

Furthermore, we showed that circuit cutting can be applied to sampling tasks.
A first sight at \cref{fig:probability_plot} may suggest only limited utility, due to a noticeable broadening of the probability distribution.
However, this impression is a consequence of the small circuit instances since \cref{overhead_sampling_task} is independent of the problem size.
As we scale to larger instances, the number of possible bit string measurement results increases exponentially with the number of qubits.
This exponential growth means that the relative impact of broadening diminishes and suggests that the apparent disadvantage of using circuit cutting for sampling tasks does not scale to larger problems.

\appendix
\subsection{QUBO Model for the TSP}
\label{QUBO Model for the TSP}
Here, we restate the QUBO model for the TSP introduced in Ref.~\cite{Lucas_2014}.
Given a TSP instance as a weighted graph $G=(V,E)$ with edge weights $d_{e}$ for $e\in E$, the model introduces a binary variable $x_{v,t}\in \{0,1\}$ for every vertex $v \in V$ and time step $t\in \{1,\dots,N\}$ where $N=|V|$.
A value of $x_{vt}=1$ indicates that vertex $v$ is visited at time step $t$ in the Hamilton cycle.
With this notion, the TSP can be modeled by
\begin{align}\label{eq:TSP}
	\min_x\ &A \sum_{v\in V}\left(1- \sum_{t=1}^{N} x_{vt}\right)^2 + A\sum_{t=1}^{N}\left(1- \sum_{v\in V} x_{vt}\right)^2 +\notag\\
 &+B\sum_{(u,v)\in V\times V}d_{uv}\sum_{i=1}^{N}x_{u,i}x_{v,i+1}\ .
\end{align}
The fist two terms penalize infeasible variable assignments whereas the last term encodes the length of the Hamilton cycle.
$A$ and $B$ are constants chosen such that any feasible solutions has smaller cost than an infeasible one.
A sufficient condition for this to hold is $A/B>N\cdot \max_{i,j} \{d_{ij}\}$.
We note that, without loss of generality, we can choose an  arbitrary vertex $\Tilde{v}\in V$ as a start vertex and fix $x_{\tilde{v},0}=1$
and $x_{v,0}=0$ for $v\in V\setminus \Tilde{v}$, reducing the number of binary variables from $N^2$ to $(N-1)^2$.

\subsection{Transformation from QUBO to MaxCut}
\label{app:qubo2maxcut}
It is well-known that any QUBO problem can be translated into an equivalent maximum cut problem  (MaxCut)~\cite{Hammer_1965,Barahona_1989,DESIMONE_1990,Juenger2021}.
Given a weighted graph $G=(V,E)$ with edge weights $w_{ij}$, where $ij \in E$, MaxCut asks for a partition of the vertices into two subsets such that the weight of connecting edges is maximized.
More formally, for a vertex subset $W\subseteq V$, we define the cut $\delta(W) \coloneqq \{ij \in E\mid i\in W, j\notin W\}$.
Furthermore, the weight of a cut $\delta(W)$ is defined as $\sum_{e\in \delta(W)}w_e$.
MaxCut asks for a cut of maximum weight.
The decision version of MaxCut is NP-complete~\cite{Karp_1972}.

A QUBO problem on $n$ variables, defined by the coefficients $\{q_{ij}|i,j\in \{1,\dots,n\}\}$, can be transformed into an equivalent MaxCut problem on $n+1$ vertices.
To this end, we consider the complete graph $K_{n+1}$ with vertices $V=\{0,1,\dots,n\}$.
For an edge $ij$ with $i,j > 0$, we define its weight as $w_{ij} = q_{ij} + q_{ji}$.
Moreover, for all edges of the form $0i$ with $i > 0$, we set $w_{0i}=\sum_{j=1}^n  q_{ij} + q_{ji}$.
Deleting edges with zero weight finally yields a weighted MaxCut instance $G=(V,E)$.
Given a cut $\delta(W)$, we construct a solution to~\cref{eq:qubo} as follows.
For $i \in \{1,\dots, n\}$, we set $x_i = 1$ if $0i\notin \delta(W)$, otherwise we set $x_i=0$.
It is easily verified that if the cut has weight $M$, the QUBO solution has value
$ Q = -M/2 + C $ where $C = 1/4\left(\sum_{e\in E}w_e + 2\sum_i q_{ii} + \sum_{i<j}q_{ij}+q_{ji} \right)$.

\subsection{MaxCut Relaxation}
\label{app:cycle_rel}
Transforming the QUBO problem into an MaxCut instance $G=(V,E)$ allows us to use established linear programming and size reduction techniques.
In particular, we use a common integer linear programming formulation of MaxCut
which is based on the observation that a cycle and a cut in $G$ coincide in an even number of edges~\cite{Barahona_1989,Juenger_2019,Rehfeldt2023}).
We introduce a binary variable $x_e\in \{0,1\}$ for all edges $e\in E$ indicating whether edge $e$ is in the cut ($x_e=1$) or not ($x_e=0$).
With this notion, MaxCut can be modeled as
\begin{subequations}
	\begin{align} 
		\max_{x} &\sum_{e \in E}w_e x_e\label{eq:Objective}\\
		\text{s.t. }&\sum_{e \in T}x_e-\sum_{e \in C\setminus T}x_e\leq |T|-1,\notag\\& \qquad\forall T \subseteq C,\ |T|\ \mathrm{odd},\ \forall C \subseteq E\ \mathrm{cycle} \label{eq:oddcycle}\\
		& 0\leq x_e\leq 1 \quad\forall e \in E \label{eq:bounds}\\
		& x_e \in \{0,1\} \quad\forall e \in E \label{eq:bin}\ .
	\end{align}
\end{subequations}
Constraints \eqref{eq:oddcycle} ensure that in any cycle the number of edges intersecting partitions is even.
This condition is necessary and sufficient for $x\in \{0,1\}^{|E|}$ to define a valid cut.
Dropping the integrality condition \eqref{eq:bin},
the model \cref{eq:Objective,eq:bounds} is called the \emph{cycle relaxation} of MaxCut.
Although the MaxCut problem is NP-hard,
optimizing the cycle relaxation can be done in polynomial time.
A key ingredient of our algorithm is, that a solution to the cycle relaxation
can be computed efficiently in practise.

\subsection{Circuit cutting and sampling}
\label{Circuit cutting and sampling}
This appendix investigates in more detail the sampling overhead arising in sampling tasks in conjunction with circuit cutting. In particular, our aim is to justify \cref{overhead_sampling_task}. To this end, we first determine the number of samples $N$ in the uncut scenario required to observe bit string $s$  with probability $1-\delta$ if the single-shot probability for $s$ is $p:=p(s)$. This probability is obtained by equating $(1-p)^N=\delta$ where the left-hand side is the probability for never observing $s$ in $N$ trials. Consequently,
\begin{equation}
    N=\Bigl \lceil\frac{\mathrm{ln}(\delta)}{\mathrm{ln}(1-p)}\Bigl \rceil\,.
\end{equation}
On the other hand, in the worst case, according to \cref{eq:sampling_task1,sampling_task2,sampling_task3}, we find $\tilde{p}=p/\kappa$ and consequently
\begin{equation}
    \tilde{N}=\Bigl \lceil\frac{\mathrm{ln}(\delta)}{\mathrm{ln}(1-p/\kappa)}\Bigl \rceil\,.
\end{equation}
In order to relate $N$ and $\tilde{N}$, we consider two separate cases. First, assume $p$ sufficiently close to zero such that $N\approx \mathrm{ln}(1/\delta)/p$ and $\tilde{N}\approx \kappa\,\mathrm{ln}(1/\delta)/p$ by Taylor expansion. Combining both expressions yields \cref{overhead_sampling_task}. Second, assume $p$ close to one. Then $N\approx1$ and $\tilde{N}\approx \mathrm{ln}(1/\delta)\kappa$ for sufficiently large $\kappa$ and $\mathrm{ln}(1/\delta)$. Once again, we approximately arrive at \cref{overhead_sampling_task}. For intermediate values of $p$, the argument still  applies and
\begin{equation}
\label{overhead_sampling_task_restated}
    \tilde{N}\approx\kappa N
\end{equation}
holds in general.

\subsection{Combining Peng et al. with Harada et al.}\label{app-combiningTwoCircuitCuttingTechniques}
\begin{figure}[ht]
    \centering
    \includegraphics[width=252pt]{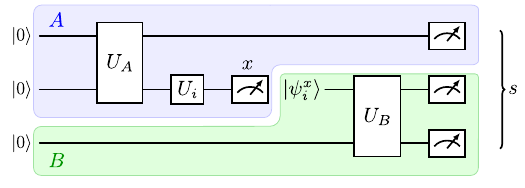}
    \caption{Schematics of a wire-cutting protocol for a single cut}
    \label{fig:circuit_single_cut}
\end{figure}

\begin{figure}[ht]
    \centering
    \includegraphics[width=252pt]{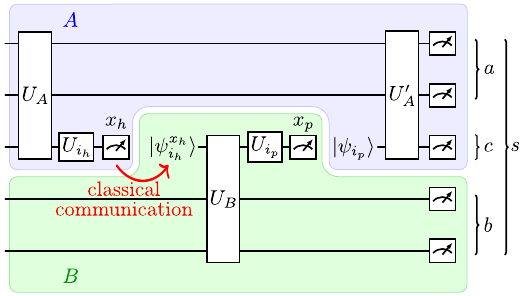}
    \caption{ Application of the combination of the methods introduced by Harada et al. and Peng et al. to the circuit layout shown. At each cut, a measurement in the computational basis rotated by $U_{i_h}$ and $ U_{i_p}$ is performed. After the cut, the qubit is re-initialized in the states $\ket{\psi_{i_h}^{x_h}}$ and $\ket{\psi_{i_p}}$. Note that only Harada's technique (first cut) uses classical communication. As a consequence, the joint protocol necessitates one-way classical communication only. }
    \label{fig:circ_harada_peng_cut}
\end{figure}

In this appendix, we detail the combination of the two circuit-cutting methods  discussed in \cref{sec:choice_cutting}. As outlined in the main text, the joint protocol leverages one-way classical communication to reduce sampling overhead, while avoiding two-way communication. We start with the general description of a wire cutting scheme for a single cut as shown in \cref{fig:circuit_single_cut}.
At the cut, the $i$th measure and prepare channel $\mathcal{F}_i$ corresponds to a measurement in the computational bases rotated by $U_i$. Depending on the outcome $x$, the state $\ket{\psi_i^x}$ is re-initialized. A protocol is said to employ classical communication if the new state explicitly depends on the measurement outcome $x$, otherwise no communication is involved.
This scheme can be extended to the QAOA circuits described in the main text with the structure shown
in \cref{fig:circ_harada_peng_cut}. The first cut is performed by the method of Harada et al. \cite{Harada2023} (classical communication), the second cut is based on the approach by Peng et al. \cite{Peng2020} (no classical communication).
Denoting the channel indices as $i_p$ ($p$ for Peng) and $i_h$ ($h$ for Harada), \cref{quasi-prob-sampling} becomes
\begin{equation}\label{equ-double-expval}
    \langle O \rangle = \sum_s\sum_{i_h, i_p} a_{i_h} a_{i_p} p(s|i_h, i_p) f(s),
\end{equation}
Next, we explicitly specify the form of the measure and prepare channels. 
For the method by Peng et al. the structure of the channels is 
\begin{align}
\label{channel1}
    \mathcal{F}_{i_p}(\cdot) &= \sum_{x_p\in \{0,1\}} c_{x_p, i_p} \mathrm{Tr} \Bigl(\;\cdot\; U_{i_p} \ket{x_p}\bra{x_p} U_{i_p}^\dagger \Bigr) \ket{\psi_{i_p}}\bra{\psi_{i_p}}\\
    \intertext{where $c_{x_p, i_p}=\pm1$. We emphasize that no classical communication is needed since $\ket{\psi_{i_p}}$ is independent of $x_p$. On the other hand, the decomposition of the identity for the method introduced by Harada et al. reads }
    \label{channel2}
    \mathcal{F}_{i_h}(\cdot) &= \sum_{x_h\in \{0,1\}} \mathrm{Tr} \Bigl(\;\cdot\; U_{i_h} \ket{x_h}\bra{x_h} U_{i_h}^\dagger \Bigr) \ket{\psi_{i_h}^{x_h}} \bra{\psi_{i_h}^{x_h}} 
\end{align}
which involves classical communication as can be seen from the index $x_h$ on $\ket{\psi_{i_h}^{x_h}}$. In both protocols the measurement and initialization basis are eigenstates of the Pauli matrices.
While the channels in \cref{channel2} are completely positive and trace preserving, and therefore represent viable physical operations, the maps in \cref{channel1} are non positive due to $c_{x_p, i_p}=-1$ for some of the measurement results. Nevertheless, such channels can be simulated by  multiplying the result with the sign of $c_{x_p, i_p}$ \cite{Mitarai2021, Brenner2023, Piveteau2022_circuitcut, Ufrecht2023}.  The decomposition according to Peng et al. comes with $\kappa_p=4$, the one introduced by Harada et al. with $\kappa_h=3$.
The probability $p(s|i_h,i_p)$ is calculated by explicitly inserting the channels \cref{channel1} and \cref{channel2}  in a circuit with structure as depicted in  \cref{fig:circ_harada_peng_cut} by which we arrive at
\begin{equation}
\label{probability for expecttion value}
    p(s|i_h, i_p) = \sum_{x_h, x_p} c_{x_p, i_p}\, p^A(a,c, x_h|i_h, i_p)\,p^B( b, x_p | x_h, i_h, i_p)\,.
\end{equation}
In this expression $p^A$ denotes the probability of observing the bit string results $a$, $c$, and $x_h$ given the channels $i_h$ and $i_p$ on the upper partition. On the other hand, $p^B$ represents  the probability of observing the results $b$ and $x_p$ on the lower partition for given channels $i_h$ and $i_p$ and measurement result $x_h$ on the upper partition. The bit string  $s$ is understood as the concatenation of $a$, $c$, and $b$.
The fact that $p^B$ is conditioned on the measurement outcome $x_h$ on the other partition, while $p^A$ is only conditioned on the applied channels, shows the necessity for one-way classical communication only.

In general, a QAOA algorithm consists of two steps, optimization of an objective based on the evaluation of an expectation value, and sampling from the trained circuit.
With \cref{probability for expecttion value} we derived the exact expression for the probability $p(s|i_h, i_p)$ in terms of $p^A$ and $p^B$. 
In contrast, in the sampling tasks we effectively generate the distribution
\begin{equation}\label{equ-ps}
    \tilde{p}(s) = \sum_{i_h, i_p} \tilde{p}(s|i_h, i_p) p(i_h,i_p)
\end{equation}
with  $ p(i_h,i_p) = |a_{i_h}a_{i_p}| /(\kappa_p\kappa_h)$.
The probability $\tilde{p}(s|i_h, i_p)$ is obtained by $c_{x_p, i_p} \rightarrow |c_{x_p, i_p}|=1$ in \cref{probability for expecttion value}. With these substitutions, we finally obtain
\begin{equation}\label{eq:p_cut}
    \tilde{p}(s) = \sum_{i_h, i_p, x_h, x_p}  \frac{|a_{i_h}a_{i_p}|}{\kappa_p\kappa_h}p^A(a,c, x_h|i_h, i_p)\,p^B( b, x_p | x_h, i_h, i_p) 
\end{equation}
the expression for the probability distribution we generate by sampling from the cut circuit.

\end{document}